\documentclass[manuscript,numberedappendix]{emulateapj}

\usepackage{amsmath}

\begin{document}

\title{X-ray Light Curve and Spectra of Shock Breakout in a Wind}
\author{Yukari Ohtani\altaffilmark{1}, Akihiro Suzuki\altaffilmark{2}, Toshikazu Shigeyama\altaffilmark{3} and Masaomi Tanaka\altaffilmark{1,4}}
\altaffiltext{1}{Center for Computational Astrophysics, National Astronomical Observatory of Japan, Mitaka, Tokyo 181-8588, Japan}
\altaffiltext{2}{Yukawa Institute for Theoretical Physics, Kyoto University, Oiwake-cho, Kitashirakawa, Sakyo-ku, Kyoto, 606-8502, Japan}
\altaffiltext{3}{Research Center for the Early Universe, Graduate School of Science, University of Tokyo, Bunkyo-ku, Tokyo 113-0033, Japan}
\altaffiltext{4}{Division of Theoretical Astronomy, National Observatory of Japan, Osawa 2-21-1, Mitaka, Tokyo 181-8588, Japan}

\begin{abstract}

We investigate the properties of X-ray emission from shock breakout of a supernova in a stellar wind.
We consider a simple model describing aspherical explosions,
in which the shock front with an ellipsoidal shape propagates into the dense circumstellar matter.
For this model, both X-ray light curves and spectra are simultaneously calculated using a Monte Carlo method.
We show that the shock breakout occurs simultaneously in all directions in a steady and spherically symmetric wind.
As a result, even for the aspherical explosion, the rise and decay timescales of the light curve do not significantly depend on the viewing angles.
This fact suggests that the light curve of the shock breakout may be used as a probe of the wind mass loss rate.
We compare our results with the observed spectrum and light curve of XRO 080109/SN 2008D.
The observation can be reproduced by an explosion with a shock velocity of 60\% of the speed of light and a circumstellar matter with a mass loss rate of $5\times10^{-4}M_{\odot}$ yr$^{-1}$.

\end{abstract}


\section{Introduction}
A core-collapse supernova emits a bright ultraviolet (UV)/X-ray flash, so-called "shock breakout", when photons generated from the shock escape upstream.
Shock breakout has been studied for several decades since \citet{1978ApJ...223L.109K} and \citet{1978ApJ...225L.133F}.
The timescale of the emission is determined by the light crossing time of the radius of the source and the diffusion timescale of photons in the unshocked or shocked matter \citep[e.g.][]{1992ApJ...393..742E,1999ApJ...510..379M}.
Shock breakout is a powerful probe of the stellar radius and the structure of the outer layer of the star, since it should be associated with all core-collapse supernovae, and the emission properties are highly sensitive to the behavior of the shock.

In 2008, the Swift/XRT accidentally detected X-ray outburst (XRO) 080109 \citep{2008Natur.453..469S}, which was associated with a type Ib supernova (SN) 2008D \citep{2008Sci...321.1185M,2009ApJ...692L..84M,2009ApJ...702..226M,2009ApJ...692.1131T}.
The luminosity rapidly reached the maximum in the first $\sim100$ sec and exponentially decayed until 600 sec  from the onset of the outburst.
The peak luminosity and total radiated energy are $6\times10^{43}$ erg s$^{-1}$ and $2\times10^{46}$ erg, respectively.
The Swift/XRT spectrum is well fitted by a power-law function, rather than a Planck function.
\citet{2008Natur.453..469S} also reported that a $UV/opt$ emission was detected by the Swift/UVOT $\sim1$ day after XRO 080109, and a decreasing X-ray emission [$L=(1.0\pm3.0)\times10^{39}$ erg s$^{-1}$ in the energy range of 0.3--10 keV]  by the Chandra X-ray Observatory $\sim10$ days after the Swift discovery.
XRO 080109 and the subsequent fainter X-ray emission is believed to originate from shock breakout and interaction of the shock with a circum-stellar matter (CSM) \citep{2008Natur.453..469S,2008ApJ...683L.135C}.

The timescale of XRO 080109 is closely related to the shock radius at the moment of breakout.
When interpreting the rise time as the light crossing time of the breakout radius,  it must be $\approx10^{12}$ cm \citep{2008Natur.453..469S}.
Since it is larger than the typical radius of a Wolf-Rayet star{,} XRO 080109 is believed to originate from a dense CSM.
The observed duration is consistent with the diffusion timescale of photons in the unshocked CSM, in which the shock breaks out at the radius of 1.1--1.6$\times10^{12}$ cm \citep{2011MNRAS.414.1715B}.
If the rise time is regarded as the shock expansion timescale, the shock radius is estimated to be $\approx 6\times10^{11}$ cm \citep{2014ApJ...788L..14S}.
Though the two estimated values are different by a factor of a few, they agree upon the excess of the breakout radius compared to the typical radius of a Wolf-Rayet progenitor.
In general, Wolf-Rayet stars blow winds with terminal velocities $v_{\rm t}$ of $v_{\rm t}\sim$1,000 km s$^{-1}$ \citep{1990ApJ...361..607P,1995A&A...299..151H} at rates $\dot{M}$ in the range of $10^{-5}$ to $10^{-4}M_{\odot}$ yr$^{-1}$ \citep{1995A&A...299..151H,1998A&A...333..956N}.
Since a wind mass loss event is known to play a significant role in the evolution of a massive star \citep{1987A&A...182..243M,1994A&AS..103...97M}, studying shock breakout also enriches the understanding of massive star evolution shortly before the explosion.
For this importance, the properties (such as timescale and luminosity evolution) of emission from the shock breakout in a wind have been predicted by several theoretical studies \citep[e.g.][]{2011MNRAS.414.1715B,2011MNRAS.415..199M,2011ApJ...729L...6C,2012ApJ...747L..17C,2012ApJ...759..108S,2014ApJ...788L..14S}.

The origin of the observed spectrum of XRO 080109 has been argued in several articles.
The observed spectrum can also be fitted by a combination of two blackbody components, but the photospheric radii are far smaller than the typical radius of a Wolf-Rayet star \citep{2008MNRAS.388..603L}.
\citet{2008Natur.453..469S} attributes the power-law spectral feature to electron ("bulk-Comptonization") scattering across a shock.
In fact, \citet{2010ApJ...719..881S} numerically examined how the photon energies increase due to this effect.
Their results imply that the observed power-law X-ray spectrum requires a shock velocity higher than $0.3c$, where $c$ denotes the speed of light.
The effect of bulk-Comptonization has also been studied by \citet{2007ApJ...664.1026W}, in which mildly relativistic shock breakout in a dense CSM is applied for the low-luminosity GRB 060218/SN 2006j \citep{2006Natur.442.1008C}.
Similar studies have been performed for shocks with lower velocities ($\lesssim10^{4}$ km s$^{-1}$) by \citet{2012ApJ...759..108S} and \citet{2012ApJ...747L..17C}, applied to the luminous Type IIn SN 2006gy.
The scattering process decreases photon energies in this particular supernova.

In addition to the presence of the CSM and the bulk-Comptonization, the asphericity of the shock front might also be important to determine the emission properties of shock breakout.
\citet{2010ApJ...717L.154S} suggested that the shape of the light curve can reflect the degree of shock asphericity and the viewing angle.
\citet{2011ApJ...727..104C} investigated the influence of shock asphericity on the light curve and spectrum by using results of their two-dimensional hydrodynamical simulations of a jet-driven supernova.
However, they do not take the influence of bulk-Compton scattering into the calculation of the spectrum.
\citet{2016ApJ...825...92S} recently performed 2D radiation hydrodynamic simulations for a blue supergiant exploding in a steady wind.
Since bipolar explosions result in the shock appearing sequentially, the light curve would have a broader peak compared to the case of a spherical shock.

Despite a large number of studies having investigated the emission properties in detail, there have been no studies that reproduce both the observed X-ray spectrum and light curve by taking bulk-Comptonization into account.
In this paper, we aim to investigate the influence of shock asphericity and bulk-Comptonization on the properties of emission from shock breakout in a wind.
For this purpose, we perform radiative transfer calculations using a Monte-Carlo method.
In Section \ref{sec:methods}, we describe the settings for the shock and Monte-Carlo calculation.
In Section \ref{sec:results}, we show the results and comparisons with the observed properties of XRO 080109/SN 2008D.
In Section \ref{sec:conclusions}, we conclude this paper.


\section{Methods}\label{sec:methods}
We calculate X-ray light curves and spectra of shock breakout emission in a dense CSM.
In the following subsections, we describe our model for the propagation of the shock (\S \ref{sec:model}) and the method to calculate radiative transfer (\S \ref{sec:radiative}).

\subsection{Model for shock}\label{sec:model}
\begin{figure}[h]
\epsscale{0.80}
\plotone{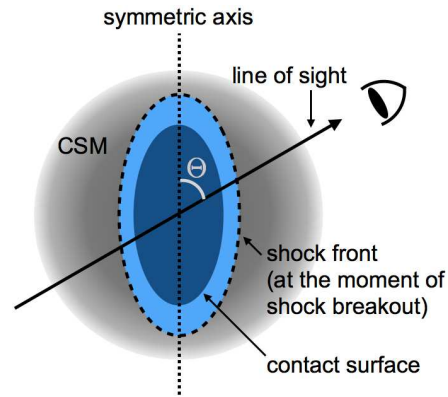}
\caption{Schematic view of an ellipsoidal shock propagating into a steady, spherically symmetric CSM. The degree of the shock asphericity is characterized by the oblateness $f$. The position of the shock front corresponds to the dashed curve at the moment of breakout.}
\label{fig:model_ellipsoid}
\end{figure}

To capture the properties of X-ray emission, we adopt a simple model of shock breakout in a wind as described below (Figure \ref{fig:model_ellipsoid}).
Our calculation does not take into account the feedback from emission to the fluid motion.

The matter is radiation dominated (the adiabatic index $\gamma$ equals to 4/3), and the radiation and matter are in thermal equilibrium below the photosphere.
Here we focus on inverse Compton scattering in the shocked CSM of interest.
For that, we think of supernova ejecta as a piston, and focus on modeling the forward shock propagating in the CSM.
We ignore the presence of the shocked ejecta.
Since the supposed ejecta density, $\approx10^{20}$ cm$^{-3}$ at the moment of breakout (hereafter $t=t_{\rm b}$ where $t$ denotes the time measured from the moment of explosion) is orders of magnitude higher than that of the shocked CSM ($\approx10^{13}$ cm$^{-3}$), photons would be absorbed or scattered immediately at the contact surface.
(Inward traveling photons generated from the shock front can be blocked by a shell filled by ejecta with a uniform density and a mass of $\Delta M_{\rm ej}\sim2\times10^{-7}M_{\odot}$, the same order of magnitude as the total mass of the shocked CSM, $M_{\rm sh}=1\times10^{-7}M_{\odot}$, at $t=t_{\rm b}$.
Here $\Delta M_{\rm ej}$ is estimated from the shell width $R_{\rm c}-r_{\rm min}$ of $1/\kappa\rho_{\rm ej}=3\times10^{4}$ cm, where $\kappa$=0.2 cm$^{2}$ g$^{-1}$ is the opacity for electron scattering, $R_{\rm c}$ the radius of the contact surface, and $\rho_{\rm ej}=1.8\times10^{-4}$ g cm$^{-3}$ the mass density of ejecta).
To see the influence of the structure behind the shock on the emission properties (shapes of the light curve and spectrum), we compare results of calculations with that using self-similar solutions of \citet{1982ApJ...258..790C} with different density structures of the ejecta. As shown in Figure \ref{fig:ar-lc-chevalier} (Appendix \ref{sec:chevalier_model}), there is no significant difference in the shapes of the X-ray light curves between the models. We obtain spectra with similar shapes as long as the density of the ejecta has a steep slope ($n\geq10$) as a function of radius (see Figure \ref{fig:ar-cpi-chevalier}).
Thus three different regions (unshocked CSM, shocked CSM, and unshocked ejecta) are under consideration. Both the shocked CSM and ejecta move at constant velocities.
The ejecta are assumed to have a uniform density and evolve in homologous expansion. The total mass is $10M_{\odot}$.

\begin{figure}[htb]
\plotone{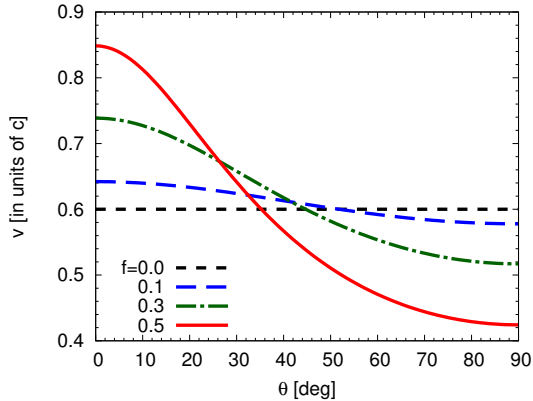}
\caption{Angular dependence of the shock velocity $v$.}
\label{fig:vej+tbreak_mid_res}
\end{figure}

We consider a shock having an ellipsoidal shape. The shock radial velocity follows the formula
\begin{eqnarray}
v(f,\theta)=\dfrac{1-f}{[(1-f)^{2}\cos^{2}\theta+\sin^{2}\theta]^{1/2}}\times v(f,0),
\end{eqnarray}
where $f$ denotes the oblateness of the shock front, $\theta$ the angle measured from the symmetric axis.
If the kinetic energy of the ejecta is fixed, the shock velocity at $\theta=0$ can be written as follows (Appendix \ref{sec:velocity_oblateness}).
\begin{eqnarray}
v(f,0)=\sqrt{3}\times v_{f=0}\times(2f^{2}-4f+3)^{-1/2},\label{eq:velocity_top}
\end{eqnarray}
where $v_{f=0}$ is the shock velocity in the spherically symmetric case.
In this study, $v_{f=0}=0.6c$.
Figure \ref{fig:vej+tbreak_mid_res} shows the angular dependence of $v$ for $f=0$ (spherical), 0.1, 0.3, and 0.5.

The wind is supposed to be stationary, spherically symmetric, and emanate from a carbon-oxygen layer.
The electron number density $n_{1}$ of the unshocked CSM follows the equation
\begin{eqnarray}
n_{1}=\dfrac{A}{r^{2}},\label{eq:n_define}
\end{eqnarray}
where $r$ is the radius measured from the center of the progenitor, and $A$ a constant.
The optical depth $\tau$ of the unshocked CSM must be equal to $c/v$ when the shock propagating at a speed $v$ breaks out.
The characteristic timescale of the emission must strongly depend on $R_{\rm b}$. We can determine the shock radius at the moment of breakout $R_{\rm b}$ for $f=0$ by a condition that the rise time of the observed emission is equal to the light crossing time $R_{\rm b}/c$.
If we adopt $\Delta t_{\rm rise}\approx100$ sec, then $R_{\rm b}$ becomes $3\times10^{12}$ cm.
Therefore the constant $A$ can be uniquely determined by the following equation,
\begin{eqnarray}
A=\frac{c}{v_{f=0}}\left[\sigma_{\rm kl}\int^{\infty}_{R_{\rm b}}r^{-2}dr\right]^{-1},\label{eq:n_factor}
\end{eqnarray}
with a free parameter $v_{f=0}$. 
$\sigma_{\rm kl}$ is the Klein-Nishina cross section for a photon with an energy corresponding to the peak energy of a blackbody radiation.
From the assumption of $v_{f=0}=0.6c$, $A=7.5\times10^{36}$ cm$^{-1}$.
When $f\neq0$, $R_{\rm b}$ has an angular dependence as written by
\begin{eqnarray}
R_{\rm b}=A\sigma_{\rm kl}\times\dfrac{v(f,\theta)}{c}.
\end{eqnarray}
This equation indicates that the shock breaks out at the same moment $R_{\rm b}/v$ in all directions (independent of $\theta$) if it has constant velocities.
If the radial velocity of the unshocked CSM is 1,000 km s$^{-1}$, the mass loss rate $\dot{M}$ becomes $5\times10^{-4}$ $M_{\odot}$ yr$^{-1}$.
The rate is one order of magnitude higher than that for an ordinary Wolf-Rayet star, but still consistent with that of a luminous blue variable \citep{1994PASP..106.1025H}.

We can estimate the thickness $\Delta R$ of the shocked CSM assuming a uniform density $n_{2}$ there.
The number density $n_{2}$ of the shocked CSM at the shock front satisfies the Rankine-Hugoniot relation
\begin{eqnarray}
\dfrac{n_{2}}{n_{1}}=\dfrac{\gamma+1}{\gamma-1}.
\label{eq:rankine}
\end{eqnarray}
A relation between the masses of the matter swept by the shock and the shocked CSM is written by
\begin{eqnarray}
4\pi R^{2}\rho_{2}\Delta R=4\pi AR,
\label{eq:shell_mass}
\end{eqnarray}
where $R=v(f,\theta)t$ is the shock radius.
From equations \eqref{eq:rankine} and \eqref{eq:shell_mass}, $\Delta R$ equals to $(\gamma-1)/(\gamma+1)\times R=R/7$. Therefore the density $\rho_{2}=2.0\times10^{-11}$ g cm$^{-3}$ and the pressure $p_{2}=6.3\times10^{9}$ g cm$^{-1}$ s$^{-2}$ at the moment of shock breakout ($R=R_{\rm b}$).
Under the assumption of radiation-dominated state and local thermodynamic equilibrium, the temperature for the shocked CSM is determined by $T_{2}=(3p_{2}/a)^{1/4}=1.3\times10^{6}$ K and that for the ejecta by $T_{\rm ej}=(3\rho_{\rm ej}v^{2}_{f=0}/a)^{1/4}=6.9\times10^{7}$ K, where $a$ is the radiation constant.
The temperature of the unshocked CSM is $T_{1}=1.0\times10^{4}$ K, which is close to the typical effective temperature of a Wolf-Rayet star \citep{2000ApJS..126..469H}.
The shocked CSM and the ejecta are assumed to have the same velocities $v(f,\theta)$.
The assumption of $10M_{\odot}$ ejecta with high velocities of $v(f,\theta)$ itself is of cause too energetic.
Again, we note that photons do not enter a deep layer of the ejecta, so that only a very low-mass ($\sim2\times10^{-7}M_{\odot}$) of ejecta is required to have high velocities as $v(f,\theta)$.
The kinetic energy of the ejecta in this region is $1\times10^{47}$ erg.
For that reason, the supposed situation is not so bad in the region calculated in this work.

\subsection{Monte-Carlo method}\label{sec:radiative}
Using the settings of Section \ref{sec:model}, we calculate radiative transfer of thermal photons by using a Monte-Carlo method.
The basic construction of the code is the same one as we used in our previous study \citep{2013ApJ...777..113O}.
Here we describe several assumptions made in this study.

Photons are isotropically generated at the shock front over a period of $\Delta t_{\rm ph}$=0.5 sec.
The  period is determined so that the total radiation energy roughly equals to the total emitted energy estimated from the Swift/XRT observation.
1,000 seed photons are generated every $5\times10^{-4}$ sec with an energy distribution following the Planck distribution in the rest frame of the fluid.
If $f=0$, the photospheric temperature is 0.11 keV (hereafter $k_{\rm B}T_{f=0}$; $T_{f=0}=T_{2}$) at the moment of breakout, and the total energy $E_{\rm tot,i}$ radiated in the time interval $\Delta t_{\rm ph}$ is $\sim6\times10^{45}$ erg.
After the shock breakout, some photons diffuse out of the shock front and reduce the pressure $p_{2}$ in the shocked CSM but not significantly change the temperature, which is proportional to $p^{1/4}_{2}$. (From the thermal energy of the shocked CSM, $4\pi R^{2}\Delta RaT^{4}_{2}=1\times10^{48}$ erg, the change in the temperature $T_{2}$ is estimated to be $\sim0.2$\%.) Thus we do not take into account this effect in the radiative transfer calculations.

We should note that the energy $E_{\rm tot,i}$ released by radiation is 1--2 orders of magnitude lower than the kinetic energy of the shocked matter, which must be comparable to the thermal radiation in an ordinary shock.
We consider that this is because a major portion of photons generated in the ejecta remains trapped.
To discuss that, a comparison of the dynamical timescale (hereafter $t_{\rm dyn}$) of the shock, $R_{\rm b}/0.6c\sim170$ sec, with the diffusion time of photons in the ejecta is needed.
We have estimated the diffusion time $t_{\rm diff}$ in a region between $r=r_{\rm min}$ and $r=R$.
If the flow expands linearly with time, the optical depth of the above region becomes unity when the shock reaches a radius (hereafter $R^{\prime\prime}_{\tau=1}$) of $\sim7\times10^{12}$ cm.
Thus the diffusion time $t_{\rm diff}$ becomes $(7-3)\times10^{12}\text{ cm}/v\sim200$ sec.
Since $t_{\rm diff}$ is longer than $t_{\rm dyn}$, it seems that a major portion of photons is still trapped in the ejecta.

The generated photons are assumed to interact with matter via inverse-Compton scattering and free-free absorption.
In the shocked CSM, the effective optical thickness $\tau_{\star}$ can be estimated by
\begin{eqnarray}
\tau_{\star}=\sqrt{\alpha^{\rm ff}(\alpha^{\rm ff}+n_{2}\sigma_{\rm kl})}\Delta R
\end{eqnarray}
where $\alpha^{\rm ff}$ is the absorption coefficient, due to free-free transition of electrons
\begin{eqnarray}
\alpha^{\rm ff}=3.7\times10^{8}T^{-1/2}Z^{2}n_{1}n_{\rm i}\nu^{-3}(1-e^{-h\nu/k_{\rm B}T})\bar{g}_{\rm ff}\text{ cm}^{-1}
\end{eqnarray}
\citep{1979rpa..book.....R}.
$T$ is the temperature, $Z$ the atomic number, $n_{\rm i}$ the number density of ion, $h$ the Planck constant, $\nu$ the frequency, and $\bar{g}_{\rm ff}(\sim1)$ the gaunt factor.
If $f=0$, $\tau_{\star}=4\times10^{-4}\ll1$ for $h\nu=$0.3 keV and $Z=8$ at the moment of shock breakout.
Therefore most photons are not absorbed by electrons.

The process of photon-electron coupling is discussed by \citet{2010ApJ...725..904N} and \citet{2010ApJ...716..781K} for shock breakout at a stellar surface, and by \citet{2012ApJ...759..108S} for that in a wind.
Here we estimate the total number of thermal photons produced by bremsstrahlung emission.
The total emissivity integrated over frequency is expressed by
\begin{eqnarray}
\varepsilon^{\rm ff}=1.4\times10^{-27}T^{1/2}n_{\rm el}n_{\rm i}Z^{2}\bar{g}_{\rm B},
\label{eq:emissivity_ff}
\end{eqnarray}
where $n_{\rm el}$ is the electron number density, $n_{\rm i}$ the ion number density, $Z$ the electric charge of the ion, and $\bar{g}_{\rm B}\sim1$ the gaunt factor \citep{1979rpa..book.....R}.
Assuming $n_{\rm el}=n_{\rm ej}$ ($n_{\rm ej}=6\times10^{19}$ cm$^{-3}$: electron number density in the ejecta), $T=T_{\rm ej}$ and fully ionized oxygen gas, Equation \eqref{eq:emissivity_ff} yields $\varepsilon^{\rm ff}=3\times10^{17}$ erg s$^{-1}$ cm$^{-3}$.
Therefore the energy $E^{\rm ff}_{\nu}$ radiated from the ejecta per unit time is roughly evaluated by $4\pi R_{\rm b}^{2}(1/\kappa\rho_{\rm ej})\varepsilon^{\rm ff}=1\times10^{48}$ erg s$^{-1}$, and the time required to release the energy of $E_{\rm tot,i}$ is $6\times10^{-3}$ sec. (Dividing $E^{\rm ff}$ by $3k_{\rm B}T_{f=0}$, we can roughly estimate the number of photons as $2\times10^{57}$ s$^{-1}$.)
Since the required time is shorter than $\Delta t_{\rm ph}$, we can consider that the generated photons are abundant enough so that the radiation and matter achieve thermal equilibrium.

Under the assumption of fully-ionized gas, possible bound-free absorption is neglected.
In order to show the validity of this assumption, we estimate the timescale for photoionization of oxygen in a process similar to that of \citet{2010ApJ...719..881S}.
The bound-free cross section of O VI ions is written by
\begin{eqnarray}
\sigma_{\rm bf}=\left(\dfrac{64\pi ng_{\rm bf}}{3\sqrt{3}Z^{2}}\right)\alpha_{\rm fs} a^{2}_{B}\left(\dfrac{\chi}{h\nu}\right)^{3}
\label{eq:cross_section_bf}
\end{eqnarray}
where $n=1$ denotes the principal quantum number, $g_{\rm bf}\sim1$ the bound-free gaunt factor, $\alpha_{\rm fs}$ the fine structure constant, $a_{B}$ the Bohr radius, and $\chi=Z^{2}\alpha^{2}_{\rm fs}m_{\rm el}c^{2}/(2n^{2})=0.87$ keV ($m_{\rm el}$: electron mass) the ionization potential \citep{1979rpa..book.....R}.
For photons with energy $3k_{\rm B}T_{f=0}$, Equation \eqref{eq:cross_section_bf} yields $\sigma_{\rm bf}=2\times10^{-18}$ cm$^{2}\gg\sigma_{\rm kl}$.
Though the fact indicates that bound-free absorption is a dominant source of opacity, the interaction would not significantly affect the non-thermal component of the X-ray spectrum due to a short timescale of photoionization.
The timescale can be estimated by the total emitted energy and number of non-thermal X-ray photons.
Using the luminosity of the thermal emission expressed by
\begin{eqnarray}
L_{\rm th}=4\pi R^{2}_{\rm b}\sigma_{\rm SB}T^{4}_{f=0}=2\times10^{46}\text{ erg s}^{-1},
\end{eqnarray}
($\sigma_{\rm SB}=ac/4$: the Stefan-Boltzmann constant) and the time interval $\Delta t_{\rm ph}$, the total energy becomes
\begin{eqnarray}
E_{\rm th}=L_{\rm th}\dfrac{c\Delta t_{\rm ph}}{v_{f=0}}=2\times10^{46}\text{ erg}.
\end{eqnarray}
Thus the number density of photons with energies of few keV is
\begin{eqnarray}
n_{\rm ph}=\dfrac{\epsilon E_{\rm th}/\text{few keV}}{4\pi R^{2}_{\rm b}c\Delta t_{\rm ph}}
\approx10^{18}\epsilon\text{ cm}^{-3},
\label{eq:density_photon}
\end{eqnarray}
where $\epsilon E_{\rm th}$ (here $\epsilon$ is supposed to be $\sim0.1$) is the total energy of non-thermal photons.
Therefore, the timescale for bound-free absorption is
\begin{eqnarray}
\tau_{\rm bf}=\dfrac{1}{c\sigma_{\rm bf}n_{\rm ph}}\approx10^{-9}\text{ s}
\end{eqnarray}

Then we estimate the timescale of radiative recombination for fully ionized oxygen written by
\begin{eqnarray}
\tau_{\rm rad}=\dfrac{1}{\alpha^{\rm rad}_{Z}\bar{n}_{2}},
\label{eq:tau_rad}
\end{eqnarray}
where
\begin{eqnarray}
\alpha^{\rm ff}_{Z}=5.197\times10^{-14}Z\beta^{-1/2}(0.4288+0.5\ln\beta+0.469\beta^{-1/3})\text{ cm}^{3}\text{ s}^{-1},
\label{eq:alpha_ff}
\end{eqnarray}
where $\beta=\chi/(k_{\rm B}T)$ \citep{1959MNRAS.119...81S}.
Substituting $T=T_{f=0}$ into the equations above, $\tau_{\rm rad}$ becomes 0.6 sec, which is far longer than $\tau_{\rm bf}$.

We can expect from a comparison of the estimated $\tau_{\rm rad}$ and $\tau_{\rm bf}$ that most elements become fully ionized immediately so that photons with energies of few keVs would not be influenced by bound-free absorption in the shocked CSM.
In the  unshocked CSM, we obtain $\tau_{\rm bf}\approx10^{-7}$ sec and $\tau_{\rm rad}=0.4$ sec, by replacing $\Delta t_{\rm ph}$ by the diffusion timescale of photons $\sim100$ sec (calculated in Section \ref{sec:res_spherical}) in Equation \eqref{eq:density_photon}, $T$ by $T_{1}$ and $\bar{n}_{2}$ by $n_{1}$ at the shock front in Equations \eqref{eq:tau_rad} and \eqref{eq:alpha_ff}.
The small $\tau_{\rm bf}/\tau_{\rm rad}$ ratio allows neglecting the influence of bound-free transition.

Once a photon reaches the surface with an optical depth $\tau$ of $10^{-2}$, it is supposed to escape from the CSM.
The calculation stops when the shock front reaches the surface of $\tau=10^{-2}$.

\section{Results}\label{sec:results}
First of all, we calculate the X-ray light curve and spectrum for a spherically symmetric supernova ($f=0$), and compare them with the observation of XRO 080109.
Then we show the dependence on the oblateness $f$ of the shock and the viewing angle $\Theta$.

\subsection{Spherically symmetric shock}\label{sec:res_spherical}
\begin{figure}
\plotone{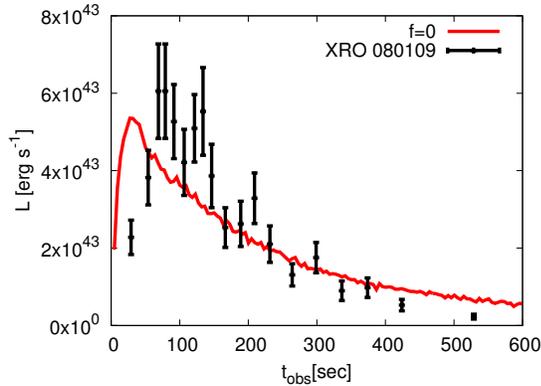}
\caption{Light curve (0.3--10 keV) of emission originating from a spherically symmetric shock. The superposed bars are the observed data of XRO 080109. The time is measured from the moment when the first photon passes a large spherical surface concentric with the ejecta.}
\label{fig:lc_f00}
\end{figure}

Figure \ref{fig:lc_f00} shows the light curve in the energy range from 0.3 keV to 10 keV covered by the Swift/XRT.
The luminosity rapidly increases for the first 40 sec, and exponentially decreases for the subsequent several hundred seconds, as $L\propto \exp[(t_{\rm obs}-t_{\rm peak})/t_{e}]$ where $t_{\rm peak}=40$ sec and $t_{\rm e}=200$ sec.
The time interval between the onset and the peak (hereafter $\Delta t_{\rm rise}$) depends primarily on the light crossing time $\Delta t_{\rm lc}$ of the size of the emerging shock, and secondarily on the diffusion timescale of photons in the shocked CSM.
We should note that there is a weak but not negligible influence of relativistic effects on the motion of photons.
A photon does not travel toward the observer if the angle between the line of sight and the radial direction exceeds $\tan^{-1}(c\gamma^{-1}v^{-1})\sim50$ deg.
Hereafter the threshold angle is referred to as $\theta_{\rm rel}$.
Therefore $\Delta t_{\rm lc}=R_{\rm b}(1-\cos\theta_{\rm rel})/c=30$ sec.
The diffusion time $t^{\prime}_{\rm diff}$ of photons in the shocked CSM is determined by the radius $R_{\tau=1}$ at which the optical depth of the shocked CSM becomes unity.
From the definition of the CSM density, Equations \eqref{eq:n_define} and \eqref{eq:n_factor}, $R_{\tau=1}\sim5\times10^{12}$ cm.
The length of $t^{\prime}_{\rm diff}$ depends on the optical depth between $r=R_{\rm b}$ and $R_{\tau=1}$. It would equal to $(R_{\rm b}-R_{\tau=1})/v=110$ sec.

\begin{figure}
\plotone{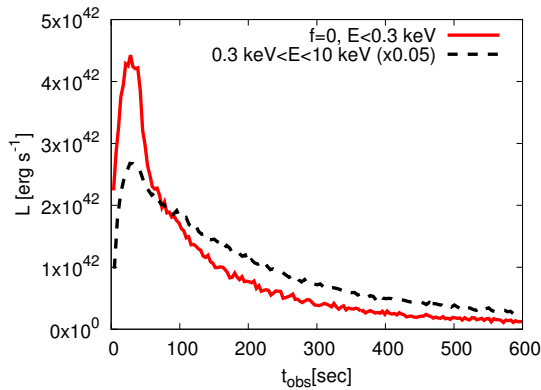}
\caption{Comparison of the light curves in the energy range of 0.3-10 keV (black dashed line) and below 0.3 keV (red solid line).}
\label{fig:lc_f00_lE}
\end{figure}

The decay time is related to the light crossing time of the radius $R_{\rm final}$ at which photons are scattered for the last time, and the photon diffusion time in the unshocked CSM.
For example, if a photon last scatters off an electron at a radius of $\sim10R_{\rm b}$, the light crossing time equals to $\sim10R_{\rm b}/c=1000$ sec.
We can expect that nearly $\sim10$\% of generated photons can scatter off electrons between this radius and infinity because the optical depth $\tau$ of the outside matter is not far smaller than 0.1, implying that the scattering probability is close to $1-\exp(-\tau)\sim0.1$.
Therefore the emission can last for several hundreds of seconds.
The time constant $t_{\rm e}$ (measured from $t=t_{\rm peak}$) is also related to the optical depth of the unshocked CSM. Because the probability that photons travel straight in the unshocked CSM is $\exp(-\tau)$ and $t_{\rm obs}-t_{\rm peak}$ is inverse proportional to $\tau$, $t_{\rm e}$ corresponds to the time when the luminosity becomes $L_{\rm peak}e^{-1}$.
The photon diffusion time is $(R_{\rm b}-R_{\tau=1})/v=110$ sec in the unshocked CSM.
The period $\Delta t_{\rm ph}$ over which the shock front emits photons does not influence on the shape of the light curve as long as it is much shorter than the duration.

The overall shape of the light curve consisting of the rapid rise and the exponential decay resembles that of the observed emission in the energy range of the Swift/XRT.
Though the length of $\Delta t_{\rm rise}$ is 40 sec in the calculation, which is shorter than the observation, it would be improved when considering a higher density CSM, in which the breakout radius $R_{\rm b}$ becomes $\sim6\times10^{12}$ cm.
Note that the maximum luminosity should not be discussed in this study, because it is proportional to the assumed period of photon generation and thus can be easily adjusted without additional simulations.
Figure \ref{fig:lc_f00_lE} shows the light curve for the energy lower than 0.3 keV.
Due to absorption, the luminosity decreases rapidly compared to that for 0.3--10 keV, so that the e-folding time $t_{\rm e}$ is as short as 120 sec.
There is no significant difference between the rise times for energy lower than 0.3 keV and for 0.3--10 keV.

\begin{figure}
\plotone{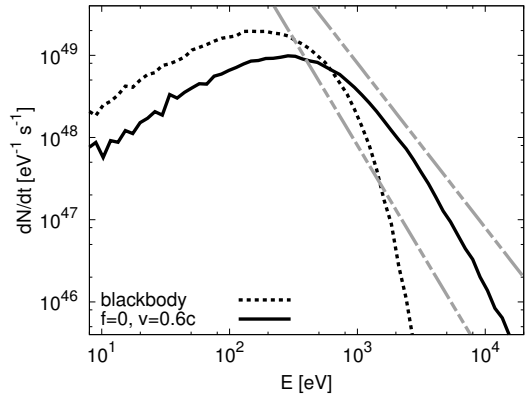}
\caption{Time-integrated spectrum of emission from a spherically symmetric shock (solid curve). The dotted curve represents a blackbody spectrum at a temperature of $k_{\rm B}T_{\rm b}$. The gradients of the straight lines are -2.0 and -2.6, which correspond to the $\pm1\sigma$ values of those of the observed Swift/XRT spectrum.}
\label{fig:spc_f00}
\end{figure}

Figure \ref{fig:spc_f00} shows the time-integrated spectrum.
The dotted line displays the blackbody with a temperature of $k_{\rm B}T_{f=0}$.
Photons following zigzag paths across the shock front receive the kinetic energies of electrons.
Their maximum energies reach $E\sim m_{\rm el}v^{2}\sim10$ keV.
The high-energy tail (1--7 keV) of the spectrum can be fitted by a power-law distribution, and the gradient lies in the $1\sigma$ error range of the observed X-ray spectrum (shown by the straight lines in Figure \ref{fig:spc_f00}).

Comparisons of our results with the observations of XRO 080109 show that the shape of the observed X-ray light curve and the spectrum can be reproduced by the emission generated from a spherically symmetric shock with a velocity of $0.6c$ and a wind with a mass loss rate of $5\times10^{-4}M_{\odot}$ yr$^{-1}$.
Here the total radiation energy $E_{\rm tot,f}$ is $2\times10^{46}$ erg, about 3 times higher than that before electron scattering ($E_{\rm tot,i}$).
The total kinetic energy $E_{\rm sh}$ of the shocked CSM is $4\pi R^{2}_{\rm b}\rho_{2}\Delta Rv^{2}=2\times10^{47}$ erg at the moment of shock breakout.
From the relatively small ratio of $E_{\rm tot,f}-E_{\rm tot,i}$ to $E_{\rm sh}$, we can expect that radiation feedback would not induce a significant change of electron temperature.

\subsection{Aspherical shock}

We investigate the influence of the asphericity of the shock on the light curve and the spectrum.
Previous calculations \citep{2010ApJ...717L.154S,2011ApJ...727..104C,2013ApJ...779...60M,2014ApJ...790...71S,2016ApJ...825...92S} investigated aspherical shock breakout in the vicinities of the stellar surfaces without thick CSM.
Caused by the significant time lags between the shocks breaking of their tops and sides, the calculated light curves show the broader peaks compared to that for a spherical shock.
Situation of our calculation is fairly different from that, as the shock breakout occurs simultaneously in all directions due to the assumptions of the thick, steady, and spherically symmetric wind and a constant shock velocity $v(f,\theta)$.
We should note that in reality, non-radial motions of ejecta play important roles along the stellar surface. If the asphericity (or "obliquity") in the ejecta motion is limited to a thin outer layer of the star and the effect of radiation is neglected, non-radial flows are believed to suppress the shock \citep{2013ApJ...779...60M}.
Thus the assumptions in our model need somewhat energetic explosion process, such as a jet-like explosion or a prolonged activity of the central engine.

\begin{figure}
\plotone{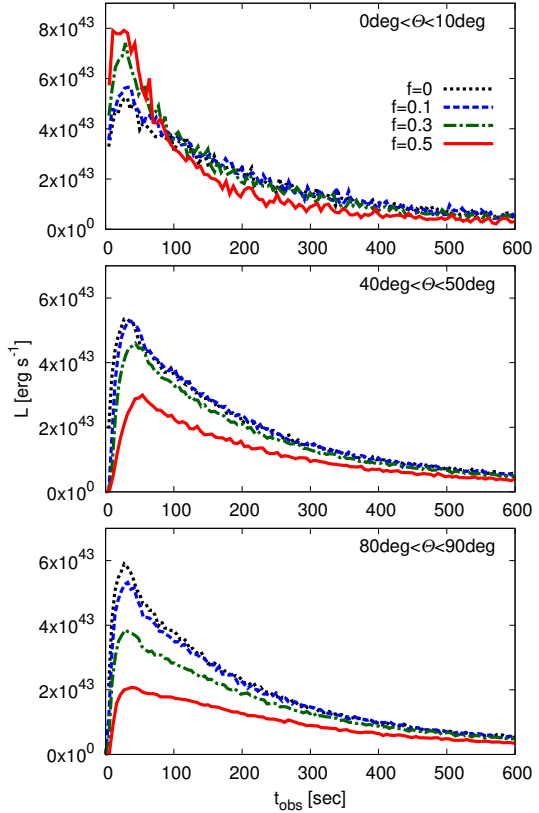}
\caption{Light curves (0.3--10 keV) of X-ray emission originating from ellipsoidal shocks. The graphs are shifted to the left side so that the times of onset corresponds to that for $f=0$.}
\label{fig:lc_sum}
\end{figure}

\begin{figure}
\plotone{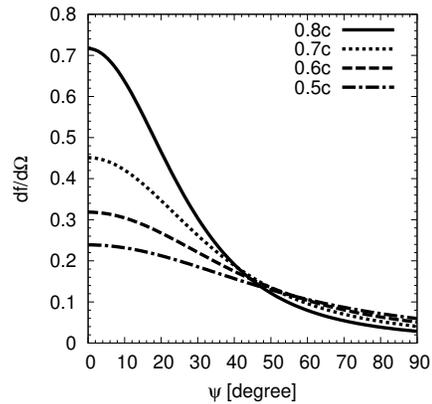}
\caption{Angular distribution of photons generated from the shock with a velocity of $0.8c$, $0.7c$, $0.6c$, $0.5c$ in the observer frame. The integral over the solid angle $\Omega$ becomes unity.}
\label{fig:beaming}
\end{figure}

Figure \ref{fig:lc_sum} shows the light curves in the energy range of the Swift/XRT when the shock has an oblateness of 0.1, 0.3, and 0.5.
The time $t_{\rm obs}$ is measured from the moment when the first photon passes a large spherical surface concentric with the ejecta.
In each panel, the flux is averaged over the angular ranges of $\theta=$[0,10], [40,50], and [80,90] deg, respectively.
The light curves show that the timescales (the rise time and the duration) of the luminosity evolution have similar values regardless of $f$ and $\Theta$.
As with the spherically symmetric case, the decay time of a few hundred sec is uniquely determined by the density distribution of the CSM. 
The reason of the similarity in the rise time is rather complicated. It depends on the shock velocity of which value varies with the inclination angle.
When the radiation intensity is concentrated in a small angle ($<\theta_{\rm rel}$), the rise time $\theta_{\rm rise}$ becomes significantly shorter than the light crossing time $\Delta t_{\rm lc}$ of the size of the emerging shock, while when the intensity is broadly distributed ($v=0.5c$ in Figure \ref{fig:beaming}), $\Delta t_{\rm rise}$ roughly equals to $\Delta t_{\rm lc}$.
For example, we can estimate the rise time observed with a viewing angle of $\Theta=0$ for a shock with an oblateness of $f=0.5$. The high shock velocity along the line of sight $\sim0.8c$ implies that most photons reaching an observer travel close to the symmetry axis ($\theta<30$ deg) due to beaming effects (see Figure \ref{fig:beaming}). Thus the rise time is approximated as $[R_{b,\theta=0}-R_{b,\theta=30^{\circ}} \cos30^{\circ}]/c\sim40$ sec where $R_{b,\theta=0}=4\times10^{12}$ cm and $R_{b,\theta=30^{\circ}}=3\times10^{12}$ cm. On the other hand, the rise time observed with a viewing angle $\Theta=90$ deg can be approximated by $\Delta t_{lc}=40$ sec because of the low shock velocity along the line of sight $\sim0.5c$.

\begin{figure}
\plotone{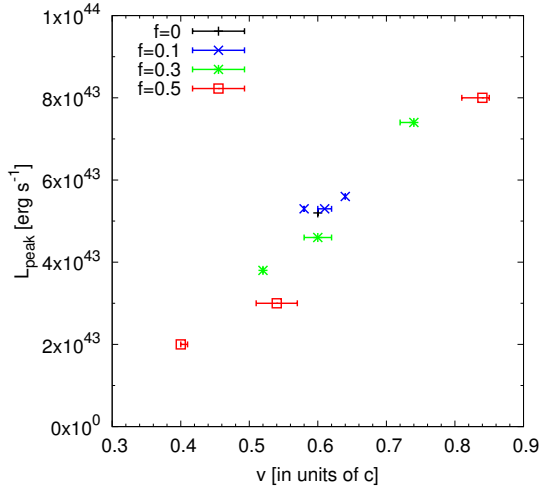}
\caption{Correction between the shock velocity and the peak luminosity. The errors shown in the horizontal bars are caused by the angular width of $\Theta$ (=[0,10], [40,50], and [80,90])}.
\label{fig:v_Lpeak}
\end{figure}

\begin{figure}
\plotone{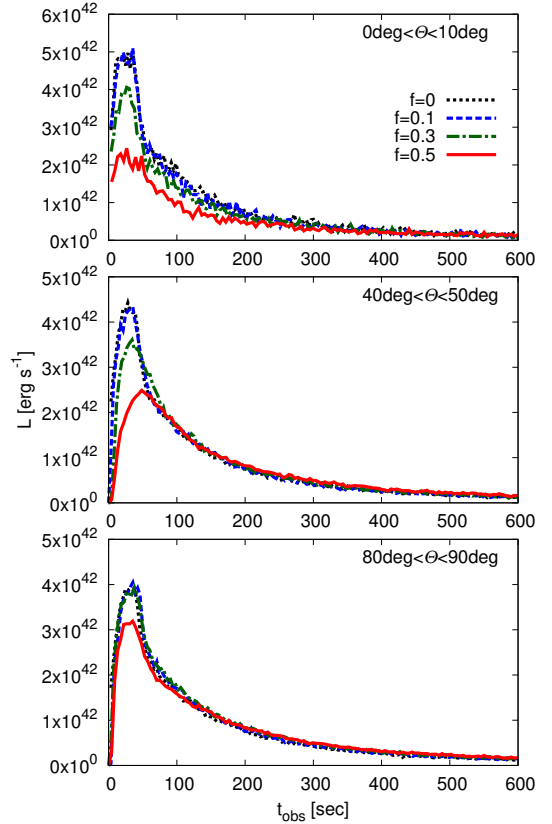}
\caption{The same as Figure \ref{fig:lc_sum} but for energies below 0.3 keV.}
\label{fig:lc_sum_lE}
\end{figure}

The peak luminosity $L_{\rm peak}$ decreases with $\Theta$, as well as the velocity of the shock propagating along the line of sight.
The relation between the shock velocity $v$ and $L_{\rm peak}$ in the energy range of 0.3--10 keV is displayed in Figure \ref{fig:v_Lpeak}. The bin widths of $v$ correspond to the ranges of the viewing angle $\Theta$. Within the range of $0.4c\lesssim v \lesssim 0.8c$, the values of $L_{\rm peak}$ tends to increase with increasing $v$ due to the bulk-Comptonization. For a lower shock velocity, such a positive correlation would become weaker because photons cannot gain much energy from electrons via electron scattering.
Figure \ref{fig:lc_sum_lE} shows the light curves for the energy lower than 0.3 keV.
Again, all of the models have similar results in the timescales and the overall evolution of the luminosity due to the simultaneous shock breakouts in all directions.
Thus the timescales responsible for the shape of the light curve becomes independent of the oblateness of the shock and the viewing angle.
Consequently, from such a light curve it would be possible to know whether the morphology of the CSM is spherical and the velocity of the shock propagating along the line of sight.
On the other hand, it would be difficult to constrain the degree of asphericities of the ejecta and shock front.


\begin{figure}
\plotone{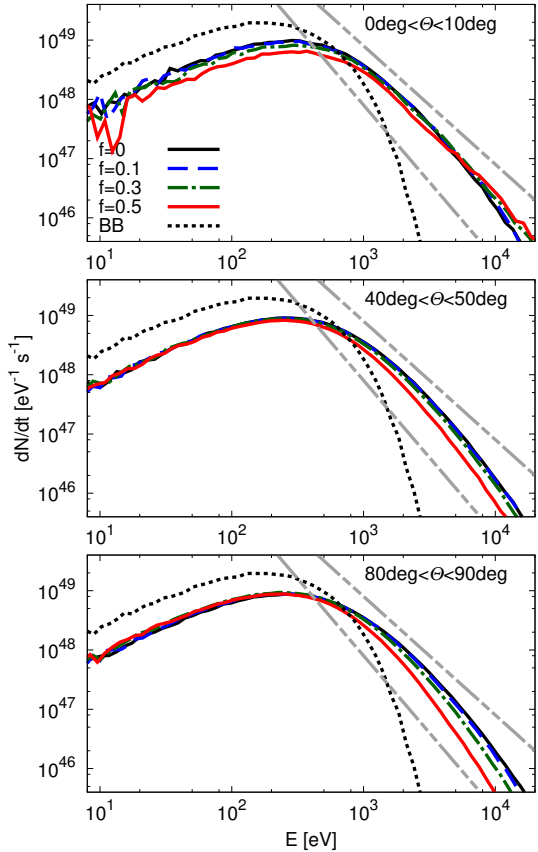}
\caption{Time-integrated spectra of X-ray emission originated from an axisymmetric shock. The dashed curve represents a blackbody spectrum at a temperature of $k_{\rm B}T_{\rm b}$. The gradients of the straight lines are -2.0 and -2.6, which correspond to the $\pm1\sigma$ values of those of the observed Swift/XRT spectrum.}
\label{fig:spc_sum}
\end{figure}

Figure \ref{fig:spc_sum} shows the time-integrated spectra for $f=$0.1, 0.3, and 0.5.
When the shock front has a finite oblateness, the power-law gradient of the high energy (1--7 keV) component is shallower along the on-axis compared to the spherically symmetric case, while it is much steeper along the off-axis.
This is because the shock has the radial velocity higher than $0.6c$ in the vicinity of the axis and lower at off-axis.
Consequently, the influence of bulk-Comptonization becomes weaker as $\Theta$ becomes larger.

\begin{figure}
\plotone{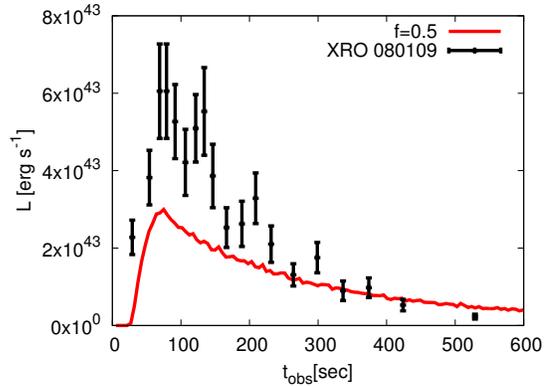}
\caption{The same as Figure \ref{fig:lc_f00} but for an ellipsoidal shock with an oblateness of $f=0.5$.}
\label{fig:lc_f05_45d}
\end{figure}

We compare the calculation for the ellipsoidal shock wave with XRO 080109.
Figure \ref{fig:lc_f05_45d} shows the light curve for $f=0.5$ and $\Theta=$45 deg as our best model.
The overall shape of the light curve is roughly consistent with observation.
In fact, an off-axis line of sight has also been suggested for XRO 080109 from late-phase observations of nebular emission lines of SN 2008D \citep{2009ApJ...700.1680T}, and our results are consistent with this interpretation.
As shown in Figure \ref{fig:spc_sum}, the high-energy spectral gradient is within a $1\sigma$ error range of the observation if the shock velocity along the line of sight is higher than $\sim0.5c$.

\section{Conclusions}\label{sec:conclusions}

We investigate the properties of X-ray emission from shock breakout in a dense CSM.
For this purpose, we calculate transfer of X-ray photons interacting with matter through Compton scattering and free-free absorption by using a Monte-Carlo method.
We also study relations between the asphericity of the shape of the ellipsoidal shock front and the observational features of the emission.

The rise time of the light curve $\Delta t_{\rm rise}$ is mainly determined by the light crossing time $\Delta t_{\rm lc}$ of the breakout radius and also slightly affected by the diffusion time $t^{\prime}_{\rm diff}$ of photons in the shocked CSM.
The major factor determining the duration of the light curve is the light crossing time of the radius at which photons
last scatter off electrons.
Even for an aspherical explosion, the properties of the light curve, such as the duration, the rise time, and the shape of the declining part, do not dramatically depend on the viewing angle as long as a steady and spherically symmetric wind is considered.
The result suggests that the characteristics of the light curve are a good probe of the CSM density or mass loss rate.

We show that both of the observed light curve and spectrum of XRO 080109/SN 2008D can be reproduced by mildly-relativistic shock breakout in a dense spherical CSM.
For a shock with the velocity of $0.6c$ and CSM with a mass loss rate of $5\times10^{-4}M_{\odot}$ yr$^{-1}$, the rise time, the duration and the shape of the calculated X-ray light curve can be consistent with the observation.
The power-law spectral gradient of the observed emission is also reproduced if the shock propagates toward the observer at a speed greater than $\sim0.5c$.

\acknowledgments
This research has been partly supported by the Grant-in-Aid for Scientific Research from JSPS (15H02075, 16H06341) and MEXT (15H00788).
Numerical computations were in part carried out on Cray XC30 at Center for Computational Astrophysics, National Astronomical Observatory of Japan.
We are grateful to Tomoya Takiwaki and Takashi Moriya for many helpful comments that improve the article. 
We also thank the Yukawa Institute for Theoretical Physics at Kyoto University. Discussions during the YITP workshop YITP-T-16-05 on "Transient Universe in the Big Survey Era: Understanding the Nature of Astrophysical Explosive Phenomena" were useful to complete this work.
Finally we thank the referee for all essential comments.

\appendix

\section{Relation of shock velocity to oblateness}\label{sec:velocity_oblateness}
The shape of the shock front should be changed under a fixed value of the explosion energy $E_{\rm kin}$.
In the following few equations, we approximate the velocity of ejecta by that of the shock front.
Here we define a typical time $t_{\rm b0}$ for each model, at which the mass density $\rho_{\rm ej}$ of ejecta has the same value as the moment of shock breakout for $f=0$.
$t_{\rm b0}$ equals $(1-f)^{-2/3}t_{\rm i}$ where $t_{\rm i}$ denotes the moment of shock breakout along the symmetric axis.
Then $E_{\rm kin}$ and the total ejecta mass $M_{\rm ej}$ are roughly expressed by the following equations.
\begin{eqnarray}
E_{\rm kin}&\approx&\iiint\dfrac{\rho_{\rm ej0}v^{2}(f,\theta)}{2}r^{2}\sin\theta drd\theta d\phi\nonumber\\
&=&\dfrac{\pi\rho_{\rm ej0}v^{2}_{0}}{3}\int^{\pi}_{0}\left[\dfrac{R^{3}(1-f)^{2}\sin\theta}{(1-f)^{2}\cos^{2}\theta+\sin^{2}\theta}\right]d\theta\nonumber\\
&=&\dfrac{2\pi\rho_{\rm ej0}v^{2}_{0}R^{3}_{0}}{9}\cdot(2f^{2}-4f+3)(1-f)^{2},\\
M_{\rm ej}&=&\dfrac{4\pi\rho_{\rm ej0}R^{3}_{0}}{3}(1-f)^{2},
\end{eqnarray}
where $\phi$ is the azimuth angle, $R$ the shock radius, and $\rho_{\rm ej0}$, $v_{0}$ and $R_{0}$ are the values at $\theta=0$.
Accordingly,
\begin{eqnarray}
E_{\rm kin}\propto v^{2}_{0}\times(2f^{2}-4f+3)
\end{eqnarray}
As a result, with a fixed $v_{f=0}=0.6c$, $v_{0}$ can be written by Equation \eqref{eq:velocity_top}.

\section{Collision of the spherically symmetric ejecta with power-law density and the circumstellar matter}\label{sec:chevalier_model}
In order to see if the assumption of uniform density significantly influences on the results of the calculation, we make a calculation for spherically symmetric ejecta with a power-law density ($\propto r^{-n}$).
The issue of the absolute luminosity is beyond the scope of this work.
We describe the hydrodynamics by using the \citeauthor{1982ApJ...258..790C} self-similar solution.
We examine the emission for $n=10$, 12 and 7 (ordinarily used to describe type Ia SNe).


\subsection{Settings}
To calculate the X-ray light curve and spectra, we describe the spherically symmetric distribution of the shocked matter and freely-expanding ejecta as follows.


The structure of the shocked CSM is determined by \citet{1963idp..book.....P}, in which the radius of the forward shock $R_{1}$ increases with time $t$ as $R_{1}\propto t^{1/\lambda_{\rm fr}}$ ($\lambda_{\rm fr}$: constant).
In the stationary CSM, if $\lambda_{\rm fr}=3/2$, the total energy of the shock becomes constant.
Here we assume the same density profile of the unshocked CSM and the shock radius $R_{1}$ at the moment of breakout as those ($n_{1}$ and $R_{{\rm b}, {f=0}}$) in Section \ref{sec:model}.

\begin{figure}
\epsscale{0.60}
\plotone{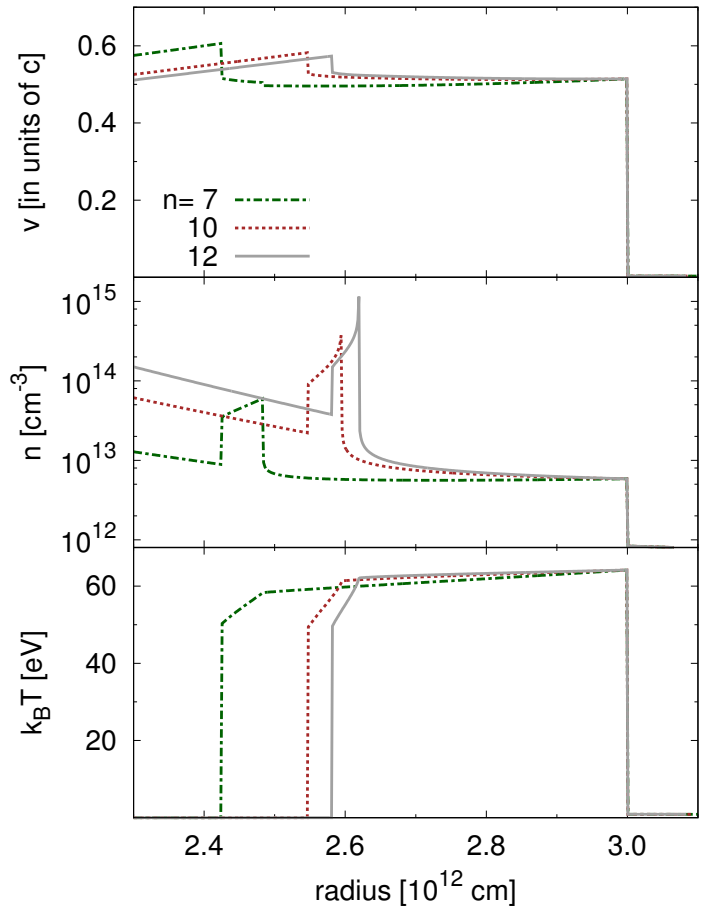}
\caption{Velocity, number density and temperature at the time of shock breakout.
The density of the unshocked CSM and ejecta is assumed to have power-law distribution.
}
\label{fig:prof-chevalier}
\end{figure}

The structure of the shocked ejecta is derived by \citet{1982ApJ...258..790C}, in which the density of the unshocked ejecta is assumed by
\begin{eqnarray}
\rho_{\rm ej}=t^{-3}\left(\dfrac{r}{tg}\right)^{-n}.
\end{eqnarray}
($g$, $n$: constant) and the radius $R_{2}\propto t^{1/\lambda_{\rm rv}}$ of the reverse shock, where $\lambda_{\rm rv}=(n-2)/(n-3)$.
Noting that the pressure and velocity of the shocked matter should be continuous at the contact surface, $g$ becomes $4.6\times10^{9}$ for $n=7$, $7.6\times10^{9}$ for $n=10$ and $9.0\times10^{9}$ for $n=12$.
Figure \ref{fig:prof-chevalier} displays the fluid profile.



Using the hydrodynamical profile above, the X-ray light curves and time-integrated spectra for the fixed shock velocity $0.6c$ at the moment of shock breakout are calculated.
The settings for the Monte-Carlo calculation is the same as those in Section \ref{sec:radiative} with the exception of the following.
If the velocity of the forward shock is fixed at $0.6c$ at the moment of shock breakout, $t_{\rm b}$ becomes 133 sec for $n=7$, 146 sec for $n=10$, and 150 sec for $n=12$.
The temperature of the matter is 64 eV (hereafter $k_{\rm B}T_{2,{\rm ch}}$) at the shock front.
From the electron number density $n_{\rm ej}$ of $\approx10^{14}$ cm$^{-3}$ in the ejecta and temperature $T$ of $T_{2,{\rm ch}}$, the total emissivity of free-free emission $\varepsilon^{\rm ff}_{\nu}$ is estimated to be $\approx10^{5}$ erg s$^{-1}$ cm$^{-3}$ and the energy $E^{\rm ff}_{\nu}$ radiated from the ejecta per unit time $\approx10^{43}$ erg s$^{-1}$.
Therefore it takes $\approx10^{2}$--$10^{3}$ sec to release the energy of $E_{\rm tot,i}$.
The fact would not lead conclusion that it is impossible to generate photons by free-free emission, but a more careful study would be needed to discuss the structure of the shock in detail.

\subsection{Light curves and spectra}
We investigate how the shapes of the light curve and X-ray spectrum differ from those in calculated in Section \ref{sec:results}.
For example, if the density of the shocked matter follows a uniform distribution, the rise time $\Delta t_{\rm rise}$ strongly depends on the light crossing time $\Delta t_{\rm lc}$ of the breakout radius and weakly on diffusion time $t^{\prime}_{\rm diff}$.
Here we investigate the dependence of the emission properties on the structure of the shock in spherically symmetric case.
%

\begin{figure}
\epsscale{0.60}
\plotone{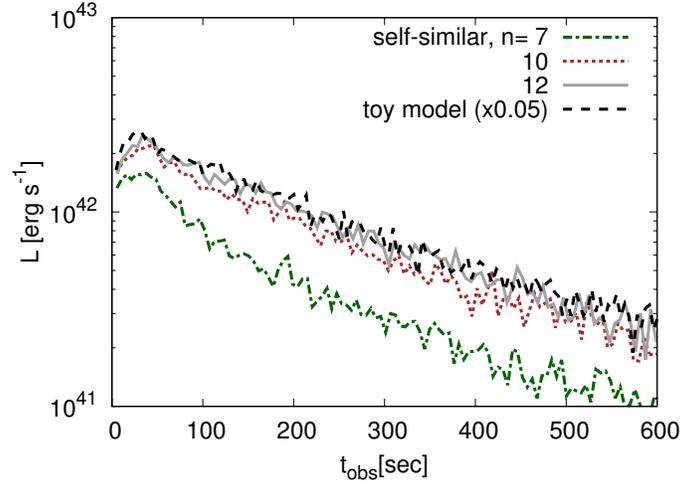}
\caption{Light curves (0.3--10 keV) calculated using the 1D \citeauthor{1982ApJ...258..790C} model. The black dashed line displays the solid line of Figure \ref{fig:lc_f00}, scaled vertically by a factor of 0.05.
}
\label{fig:ar-lc-chevalier}
\end{figure}

Figure \ref{fig:ar-lc-chevalier} shows the resultant light curves in the energy range of 0.3--10 keV.
The graphs are compared with the model with uniform density distribution (black dotted line; the luminosity is scaled by a factor of 0.05).
If $n=10$ and 12, the overall shapes of the light curves are quite similar to that for the model with uniform density.
In order to know on what timescales the rise time $\Delta t_{\rm rise}$ and duration depend, we compare the diffusion time $t^{\prime}_{\rm diff}=(R^{\prime\prime}_{\tau=1}-r_{\rm min})/v$ in the shocked matter and unshocked CSM, and the light crossing time $\Delta t_{\rm lc}$ of the size of the emerging shock. 
Here $R^{\prime\prime}_{\tau=1}\sim7\times10^{12}$ cm.
From the radius $r_{\rm min}$ (which satisfies $\int^{R_{\rm c}}_{\rm rmin}\kappa\rho dr$=1), $2.58\times10^{12}$ cm for $n=10$ and $2.61\times10^{12}$ cm  for $n=12$, $t^{\prime}_{\rm diff}$ is estimated to be $\sim(R^{\prime\prime}_{\tau=1}-r_{\rm min})/0.5c=300$ sec, which is longer than the light crossing $\Delta t_{\rm lc}$ estimated in Section \ref{sec:res_spherical}.
For this reason, $\Delta t_{\rm lc}$ can be said to be the primary factor in determining $\Delta t_{\rm rise}$ rather than $t^{\prime}_{\rm diff}$.

Though a model with $n=7$ is fainter than those with $n=10$ and $n=12$ due to the low temperature of the shocked matter, the timescale of the brightening and the duration are not significantly different from those with $n=10$ and 12.
(The radius $r_{\rm min}$ equals $2.45\times10^{12}$ cm so that $t^{\prime}_{\rm diff}>\Delta t_{\rm lc}$.)

\begin{figure}
\epsscale{0.60}
\plotone{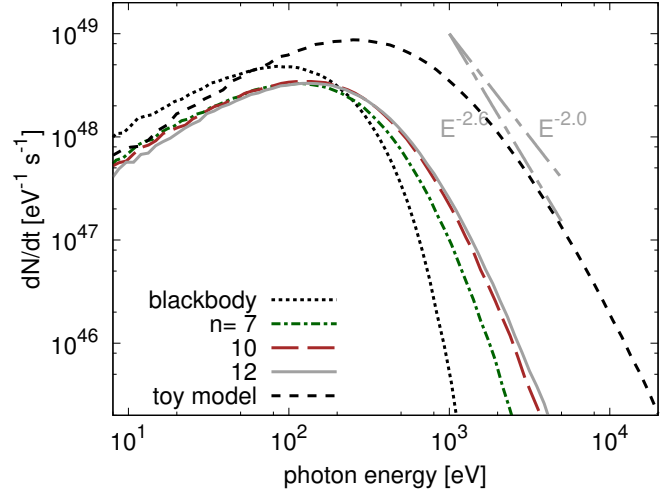}
\caption{Time-integrated spectra calculated using the 1D \citeauthor{1982ApJ...258..790C} self-similar solution for $n=7$, 10 and 12, and that of a blackbody at a temperature of $k_{\rm B}T_{2,{\rm ch}}$ (black dotted line). The black short-dashed line displays the solid curve of Figure \ref{fig:spc_f00}.}
\label{fig:ar-cpi-chevalier}
\end{figure}

Figure \ref{fig:ar-cpi-chevalier} shows the time-integrated spectrum.
Here we examine how the spectral gradient of the high-energy (1--7 keV) component changes with the motion of the matter.
In comparison with the model with the uniform density and velocity distribution (black short-dotted line), the spectral gradient is steeper if the structure behind the shock is considered.
It is because the velocity of the shocked matter is lower than the shock wave.
For Figure \ref{fig:ar-cpi-chevalier}, we can say that the spectral gradient of the high-energy tail is determined by the effect of bulk-Comptonization, so it can be a source of information on the shock velocity.

\bibliographystyle{apj}
\bibliography{hogehoge}

\begin{thebibliography}{40}
\expandafter\ifx\csname natexlab\endcsname\relax\def\natexlab#1{#1}\fi

\bibitem[{{Balberg} \& {Loeb}(2011)}]{2011MNRAS.414.1715B}
{Balberg}, S., \& {Loeb}, A. 2011, \mnras, 414, 1715

\bibitem[{{Campana} {et~al.}(2006){Campana}, {Mangano}, {Blustin}, {Brown},
  {Burrows}, {Chincarini}, {Cummings}, {Cusumano}, {Della Valle}, {Malesani},
  {M{\'e}sz{\'a}ros}, {Nousek}, {Page}, {Sakamoto}, {Waxman}, {Zhang}, {Dai},
  {Gehrels}, {Immler}, {Marshall}, {Mason}, {Moretti}, {O'Brien}, {Osborne},
  {Page}, {Romano}, {Roming}, {Tagliaferri}, {Cominsky}, {Giommi}, {Godet},
  {Kennea}, {Krimm}, {Angelini}, {Barthelmy}, {Boyd}, {Palmer}, {Wells}, \&
  {White}}]{2006Natur.442.1008C}
{Campana}, S., {Mangano}, V., {Blustin}, A.~J., {et~al.} 2006, \nat, 442, 1008

\bibitem[{{Chevalier}(1982)}]{1982ApJ...258..790C}
{Chevalier}, R.~A. 1982, \apj, 258, 790

\bibitem[{{Chevalier} \& {Fransson}(2008)}]{2008ApJ...683L.135C}
{Chevalier}, R.~A., \& {Fransson}, C. 2008, \apjl, 683, L135

\bibitem[{{Chevalier} \& {Irwin}(2011)}]{2011ApJ...729L...6C}
{Chevalier}, R.~A., \& {Irwin}, C.~M. 2011, \apjl, 729, L6

\bibitem[{{Chevalier} \& {Irwin}(2012)}]{2012ApJ...747L..17C}
---. 2012, \apjl, 747, L17

\bibitem[{{Couch} {et~al.}(2011){Couch}, {Pooley}, {Wheeler}, \&
  {Milosavljevi{\'c}}}]{2011ApJ...727..104C}
{Couch}, S.~M., {Pooley}, D., {Wheeler}, J.~C., \& {Milosavljevi{\'c}}, M.
  2011, \apj, 727, 104

\bibitem[{{Ensman} \& {Burrows}(1992)}]{1992ApJ...393..742E}
{Ensman}, L., \& {Burrows}, A. 1992, \apj, 393, 742

\bibitem[{{Falk}(1978)}]{1978ApJ...225L.133F}
{Falk}, S.~W. 1978, \apjl, 225, L133

\bibitem[{{Hamann} {et~al.}(1995){Hamann}, {Koesterke}, \&
  {Wessolowski}}]{1995A&A...299..151H}
{Hamann}, W.-R., {Koesterke}, L., \& {Wessolowski}, U. 1995, \aap, 299, 151

\bibitem[{{Herald} {et~al.}(2000){Herald}, {Schulte-Ladbeck}, {Eenens}, \&
  {Morris}}]{2000ApJS..126..469H}
{Herald}, J.~E., {Schulte-Ladbeck}, R.~E., {Eenens}, P.~R.~J., \& {Morris}, P.
  2000, \apjs, 126, 469

\bibitem[{{Humphreys} \& {Davidson}(1994)}]{1994PASP..106.1025H}
{Humphreys}, R.~M., \& {Davidson}, K. 1994, \pasp, 106, 1025

\bibitem[{{Katz} {et~al.}(2010){Katz}, {Budnik}, \&
  {Waxman}}]{2010ApJ...716..781K}
{Katz}, B., {Budnik}, R., \& {Waxman}, E. 2010, \apj, 716, 781

\bibitem[{{Klein} \& {Chevalier}(1978)}]{1978ApJ...223L.109K}
{Klein}, R.~I., \& {Chevalier}, R.~A. 1978, \apjl, 223, L109

\bibitem[{{Li}(2008)}]{2008MNRAS.388..603L}
{Li}, L.-X. 2008, \mnras, 388, 603

\bibitem[{{Maeder} \& {Meynet}(1987)}]{1987A&A...182..243M}
{Maeder}, A., \& {Meynet}, G. 1987, \aap, 182, 243

\bibitem[{{Malesani} {et~al.}(2009){Malesani}, {Fynbo}, {Hjorth}, {Leloudas},
  {Sollerman}, {Stritzinger}, {Vreeswijk}, {Watson}, {Gorosabel},
  {Micha{\l}owski}, {Th{\"o}ne}, {Augusteijn}, {Bersier}, {Jakobsson},
  {Jaunsen}, {Ledoux}, {Levan}, {Milvang-Jensen}, {Rol}, {Tanvir}, {Wiersema},
  {Xu}, {Albert}, {Bayliss}, {Gall}, {Grove}, {Koester}, {Leitet}, {Pursimo},
  \& {Skillen}}]{2009ApJ...692L..84M}
{Malesani}, D., {Fynbo}, J.~P.~U., {Hjorth}, J., {et~al.} 2009, \apjl, 692, L84

\bibitem[{{Matzner} {et~al.}(2013){Matzner}, {Levin}, \&
  {Ro}}]{2013ApJ...779...60M}
{Matzner}, C.~D., {Levin}, Y., \& {Ro}, S. 2013, \apj, 779, 60

\bibitem[{{Matzner} \& {McKee}(1999)}]{1999ApJ...510..379M}
{Matzner}, C.~D., \& {McKee}, C.~F. 1999, \apj, 510, 379

\bibitem[{{Mazzali} {et~al.}(2008){Mazzali}, {Valenti}, {Della Valle},
  {Chincarini}, {Sauer}, {Benetti}, {Pian}, {Piran}, {D'Elia}, {Elias-Rosa},
  {Margutti}, {Pasotti}, {Antonelli}, {Bufano}, {Campana}, {Cappellaro},
  {Covino}, {D'Avanzo}, {Fiore}, {Fugazza}, {Gilmozzi}, {Hunter}, {Maguire},
  {Maiorano}, {Marziani}, {Masetti}, {Mirabel}, {Navasardyan}, {Nomoto},
  {Palazzi}, {Pastorello}, {Panagia}, {Pellizza}, {Sari}, {Smartt},
  {Tagliaferri}, {Tanaka}, {Taubenberger}, {Tominaga}, {Trundle}, \&
  {Turatto}}]{2008Sci...321.1185M}
{Mazzali}, P.~A., {Valenti}, S., {Della Valle}, M., {et~al.} 2008, Science,
  321, 1185

\bibitem[{{Meynet} {et~al.}(1994){Meynet}, {Maeder}, {Schaller}, {Schaerer}, \&
  {Charbonnel}}]{1994A&AS..103...97M}
{Meynet}, G., {Maeder}, A., {Schaller}, G., {Schaerer}, D., \& {Charbonnel}, C.
  1994, \aaps, 103

\bibitem[{{Modjaz} {et~al.}(2009){Modjaz}, {Li}, {Butler}, {Chornock},
  {Perley}, {Blondin}, {Bloom}, {Filippenko}, {Kirshner}, {Kocevski},
  {Poznanski}, {Hicken}, {Foley}, {Stringfellow}, {Berlind}, {Barrado y
  Navascues}, {Blake}, {Bouy}, {Brown}, {Challis}, {Chen}, {de Vries},
  {Dufour}, {Falco}, {Friedman}, {Ganeshalingam}, {Garnavich}, {Holden},
  {Illingworth}, {Lee}, {Liebert}, {Marion}, {Olivier}, {Prochaska},
  {Silverman}, {Smith}, {Starr}, {Steele}, {Stockton}, {Williams}, \&
  {Wood-Vasey}}]{2009ApJ...702..226M}
{Modjaz}, M., {Li}, W., {Butler}, N., {et~al.} 2009, \apj, 702, 226

\bibitem[{{Moriya} {et~al.}(2011){Moriya}, {Tominaga}, {Blinnikov}, {Baklanov},
  \& {Sorokina}}]{2011MNRAS.415..199M}
{Moriya}, T., {Tominaga}, N., {Blinnikov}, S.~I., {Baklanov}, P.~V., \&
  {Sorokina}, E.~I. 2011, \mnras, 415, 199

\bibitem[{{Nakar} \& {Sari}(2010)}]{2010ApJ...725..904N}
{Nakar}, E., \& {Sari}, R. 2010, \apj, 725, 904

\bibitem[{{Nugis} {et~al.}(1998){Nugis}, {Crowther}, \&
  {Willis}}]{1998A&A...333..956N}
{Nugis}, T., {Crowther}, P.~A., \& {Willis}, A.~J. 1998, \aap, 333, 956

\bibitem[{{Ohtani} {et~al.}(2013){Ohtani}, {Suzuki}, \&
  {Shigeyama}}]{2013ApJ...777..113O}
{Ohtani}, Y., {Suzuki}, A., \& {Shigeyama}, T. 2013, \apj, 777, 113

\bibitem[{{Parker}(1963)}]{1963idp..book.....P}
{Parker}, E.~N. 1963,

\bibitem[{{Prinja} {et~al.}(1990){Prinja}, {Barlow}, \&
  {Howarth}}]{1990ApJ...361..607P}
{Prinja}, R.~K., {Barlow}, M.~J., \& {Howarth}, I.~D. 1990, \apj, 361, 607

\bibitem[{{Rybicki} \& {Lightman}(1979)}]{1979rpa..book.....R}
{Rybicki}, G.~B., \& {Lightman}, A.~P. 1979, {Radiative processes in
  astrophysics}

\bibitem[{{Salbi} {et~al.}(2014){Salbi}, {Matzner}, {Ro}, \&
  {Levin}}]{2014ApJ...790...71S}
{Salbi}, P., {Matzner}, C.~D., {Ro}, S., \& {Levin}, Y. 2014, \apj, 790, 71

\bibitem[{{Seaton}(1959)}]{1959MNRAS.119...81S}
{Seaton}, M.~J. 1959, \mnras, 119, 81

\bibitem[{{Soderberg} {et~al.}(2008){Soderberg}, {Berger}, {Page}, {Schady},
  {Parrent}, {Pooley}, {Wang}, {Ofek}, {Cucchiara}, {Rau}, {Waxman}, {Simon},
  {Bock}, {Milne}, {Page}, {Barentine}, {Barthelmy}, {Beardmore}, {Bietenholz},
  {Brown}, {Burrows}, {Burrows}, {Byrngelson}, {Cenko}, {Chandra}, {Cummings},
  {Fox}, {Gal-Yam}, {Gehrels}, {Immler}, {Kasliwal}, {Kong}, {Krimm},
  {Kulkarni}, {Maccarone}, {M{\'e}sz{\'a}ros}, {Nakar}, {O'Brien}, {Overzier},
  {de Pasquale}, {Racusin}, {Rea}, \& {York}}]{2008Natur.453..469S}
{Soderberg}, A.~M., {Berger}, E., {Page}, K.~L., {et~al.} 2008, \nat, 453, 469

\bibitem[{{Suzuki} {et~al.}(2016){Suzuki}, {Maeda}, \&
  {Shigeyama}}]{2016ApJ...825...92S}
{Suzuki}, A., {Maeda}, K., \& {Shigeyama}, T. 2016, \apj, 825, 92

\bibitem[{{Suzuki} \& {Shigeyama}(2010{\natexlab{a}})}]{2010ApJ...719..881S}
{Suzuki}, A., \& {Shigeyama}, T. 2010{\natexlab{a}}, \apj, 719, 881

\bibitem[{{Suzuki} \& {Shigeyama}(2010{\natexlab{b}})}]{2010ApJ...717L.154S}
---. 2010{\natexlab{b}}, \apjl, 717, L154

\bibitem[{{Svirski} \& {Nakar}(2014)}]{2014ApJ...788L..14S}
{Svirski}, G., \& {Nakar}, E. 2014, \apjl, 788, L14

\bibitem[{{Svirski} {et~al.}(2012){Svirski}, {Nakar}, \&
  {Sari}}]{2012ApJ...759..108S}
{Svirski}, G., {Nakar}, E., \& {Sari}, R. 2012, \apj, 759, 108

\bibitem[{{Tanaka} {et~al.}(2009{\natexlab{a}}){Tanaka}, {Yamanaka}, {Maeda},
  {Kawabata}, {Hattori}, {Minezaki}, {Valenti}, {Della Valle}, {Sahu},
  {Anupama}, {Tominaga}, {Nomoto}, {Mazzali}, \& {Pian}}]{2009ApJ...700.1680T}
{Tanaka}, M., {Yamanaka}, M., {Maeda}, K., {et~al.} 2009{\natexlab{a}}, \apj,
  700, 1680

\bibitem[{{Tanaka} {et~al.}(2009{\natexlab{b}}){Tanaka}, {Tominaga}, {Nomoto},
  {Valenti}, {Sahu}, {Minezaki}, {Yoshii}, {Yoshida}, {Anupama}, {Benetti},
  {Chincarini}, {Della Valle}, {Mazzali}, \& {Pian}}]{2009ApJ...692.1131T}
{Tanaka}, M., {Tominaga}, N., {Nomoto}, K., {et~al.} 2009{\natexlab{b}}, \apj,
  692, 1131

\bibitem[{{Wang} {et~al.}(2007){Wang}, {Li}, {Waxman}, \&
  {M{\'e}sz{\'a}ros}}]{2007ApJ...664.1026W}
{Wang}, X.-Y., {Li}, Z., {Waxman}, E., \& {M{\'e}sz{\'a}ros}, P. 2007, \apj,
  664, 1026

\end{thebibliography}

\end{document}